\documentstyle[12pt]{article}
\begin{document}
\title{Stability of Matter in Magnetic Fields}
\author{Elliott~H.~Lieb$^{1,2}$, Michael Loss$^3$ and Jan Philip Solovej$^2$\\
\footnotesize \it $^1$Department of Physics, Jadwin Hall, Princeton University,
P.~O.~Box 708, Princeton, New Jersey 08544\\ \footnotesize \it
$^2$Department of Mathematics, Fine Hall, Princeton University,
Princeton, New Jersey 08544\\ \footnotesize \it
$^3$School of Mathematics, Georgia Institute of Technology, Atlanta,
Georgia 30332}
\date{April 11 (revised June 11)}
\maketitle

\begin{abstract}
In the presence of arbitrarily large magnetic fields, matter
composed of electrons and nuclei was known to be unstable if $\alpha$
or $Z$ is too large. Here we prove that matter
{\it is stable\/} if $\alpha<0.06$ and $Z\alpha^2<0.04$.
\end{abstract}

\renewcommand{\baselinestretch}{1.5}

One of the remaining unsolved problems connected with the stability of
matter is the
inclusion of arbitrary magnetic fields.
The model is a caricature of QED which invites
speculations about stability of QED for large fine structure constant,
$\alpha$, but that is not our focus here and
we refer to \cite{FLL} for a discussion of these and related matters.
The Hamiltonian for $N$
electrons and $K$ fixed
nuclei of charge $Ze$ with magnetic field $\mbox{\boldmath $B$}(x)=
\nabla \times \mbox{\boldmath $A$}(x)$, including the
field energy, $\varepsilon\int B^2$, is
\begin{equation}\label{eq:1}
	H=\sum_{i=1}^N {\cal T}_i +
	V_{\rm c} +
	\varepsilon\int B(x)^2 {\rm d} ^3x \ ,
\end{equation}
where ${\cal T}\equiv [\mbox{\boldmath $\sigma$}
\cdot(\mbox{\boldmath $p$}+\mbox{\boldmath $A$})]^2=
(\mbox{\boldmath $p$}+\mbox{\boldmath $A$})^2+
\mbox{\boldmath $\sigma$} \cdot \mbox{\boldmath $B$}$ is the
Pauli operator.  The Coulomb energy is
\begin{eqnarray}
V_{\rm c}&=&- Z\sum_{i=1}^N\sum_{j=1}^K |x_i-R_j|^{-1} +
\sum_{1 \le i<j \le N} |x_i-x_j|^{-1} \nonumber \\
&&+ Z^2 \sum_{1 \le i<j \le K}|R_i-R_j|^{-1} \label{eq:2}\ ,
\end{eqnarray}
with $R_j$ being the coordinates of the nuclei and $x_i$ the
electron coordinates.
The energy unit is 4 Rydbergs $=2mc^2\alpha^2$, $\alpha=e^2/\hbar c$,
length unit $=\hbar^2/2me^2$
and $\varepsilon=(8\pi\alpha^2)^{-1}$.
Notice that $\alpha$ appears in (\ref{eq:1}) only through $\varepsilon$.

 The negative particles, i.e., the electrons, are necessarily
spin 1/2 fermions which, for generality, we assume to exist
in $q/2$ flavors (e.g., $q=6$). The ground state energy is denoted
by $E$.

Starting with the 1967 pioneering work of Dyson and Lenard
we now understand stability for arbitrarily many electrons
and nuclei, with $\mbox{\boldmath $B$}=0$,
in the context of the nonrelativistic Schr\"odinger
equation. Later it was extended to the ``relativistic''
Schr\"odinger equation in which $\mbox{\boldmath $p$}^2/{2m}$
is replaced by $(c^2\mbox{\boldmath $p$}^2+m^2c^4)^{1/2}$
(see \cite{L1} for a review).
These proofs also hold with the inclusion of a magnetic
field coupled to the {\it orbital motion} of the electrons, i.e.,
$\mbox{\boldmath $p$} \rightarrow \mbox{\boldmath $p$}+
\mbox{\boldmath $A$}$,  but no Zeeman {\boldmath $\sigma\cdot B$}
term.

Stability of matter has two meanings: (1) $E$ is finite for
arbitrary N and K;
(2) $E \ge -C_1(N+K)$ for some constant $C_1$ independent of $N$, $K$
and $R_j$.
In the nonrelativistic case (2) holds. In the relativistic
case, (1) implies (2) but (1) requires two conditions:
$Z\alpha \le C_2$ and $\alpha \le C_3$ with $C_2$ and $C_3$
being universal constants, the best available values being
in \cite{LY}, Theorems 1 and 2.
The inclusion of $\mbox{\boldmath $B$}$ changes
$E$, but the point is that while $C_1, C_2, C_3$ depend on
$q$, they can be chosen to be independent of $\mbox{\boldmath $B$}$.

The situation changes dramatically when the magnetic moments of the
electrons are allowed to interact with the magnetic field via the
$\mbox{\boldmath $\sigma$} \cdot \mbox{\boldmath $B$}$
term, as in (\ref{eq:1}). The reason for this is simple:
The Pauli operator ${\cal T}$ is nonnegative but it is much weaker than
$(\mbox{\boldmath $p$}+\mbox{\boldmath $A$})^2$. Indeed, it can even
have square integrable zero-modes
\cite{LoY}, ${\cal T}\psi=0$, for suitable $\mbox{\boldmath $A$}(x)$,
which cause instability for large $Z\alpha^2$.

It is known \cite{AHS} that without the field energy term,
$\varepsilon\int B^2$,
in (\ref{eq:1}) arbitrarily large $\mbox{\boldmath $B$}$
fields can cause arbitrarily negative
energies, $E$, even for hydrogen. The field energy, hopefully,
stabilizes the situation, and our goal is to show that $E$ is finite
for (\ref{eq:1}),
{\it even after minimizing over all possible $\mbox{\boldmath $B$}$ fields}
and all possible $R_j$.

One of our results on magnetic stability is:

{\it Theorem~1:  The ground state energy of H
satisfies
\begin{equation}\label{eq:3}
	E \ge -2.6\,q^{2/3}\max\{Q(Z)^2,Q(5.7q)^2\}N^{1/3}K^{2/3} \ ,
\end{equation}
with $Q(t)\equiv t+\sqrt{2t}+2.2$, provided that
\begin{equation}\label{eq:4}
	qZ\alpha^2 \le 0.082\quad  \mbox{and}\quad   q\alpha \le 0.12\ .
\end{equation}
}

In (\ref{eq:2}) all the nuclear charges are set equal to $Z$. As far as
stability is concerned this is no restriction \cite{DL} since the
energy is concave in each charge $Z_j$ and hence stability holds in the
``cube'' $\{0 \le Z_j \le Z\}_{j=1}^K$ if it holds when all $Z_j=Z$. It
also follows from this that $E$ is a decreasing function of $Z$.
Moreover, since $\varepsilon \propto \alpha^{-2}$ it follows that $E$
is a decreasing function of $\alpha$, a fact that will be important
later. The
form of (\ref{eq:3}) is the best possible for $Z\geq 1$, as we know from other
studies \cite{L1}.

Our actual condition for stability given after (\ref{eq:c8}) is rather
complicated, but very much more general than (\ref{eq:4})---which is
only representative.
The results after (\ref{eq:c8}) show, e.g., that when $\alpha=1/137$, and
$q=2$, $Z$ can be as large as 1050.
The large values of $Z$ and $\alpha$ are important because the
comfortable distance of the critical values
from the physical values $Z\leq 92$, $\alpha=1/137$ implies that the
effect studied here is merely a small perturbation.

Our proof of Theorem~1 will require a new technique---a running
energy-scale renormalization of ${\cal T}$. A by-product of this
is a Lieb-Thirring type inequality for ${\cal T}$:

{\it Theorem~2: If $\varepsilon_1\leq\varepsilon_2,\ldots<0$ are
the negative eigenvalues of ${\cal T}-U$, for a potential $-U(x)\leq0$
then
\begin{eqnarray}
        \sum|\varepsilon_i|&\leq&
        a_\gamma\int U(x)^{5/2}{\rm d}^3x\nonumber\\
        &+&b_\gamma\left(\int B(x)^2{\rm d}^3 x\right)^{3/4}
        \left(\int U(x)^4{\rm d}^3x\right)^{1/4} \label{eq:lt2}
\end{eqnarray}
for all $0<\gamma<1$, where
$a_\gamma=(2^{3/2}/5)(1-\gamma)^{-1}L_3$ and $b_\gamma=
3^{1/4} 2^{-9/4}\pi \gamma^{-3/8}(1-\gamma)^{-5/8}L_3$.
We can take $L_3$, defined below, to be 0.1156.}

More generally the second term in (\ref{eq:lt2}) can be replaced by
$(\int B^{3q/2})^{1/q}(\int U^p)^{1/p}$, where
$p^{-1}+q^{-1}=1$.

The investigation of this problem started in \cite{FLL,LL} where type 2
stability was proved (for suitable $Z$,$\alpha$ and $q=2$) for $K=1$
and arbitrary $N$ (if $(Z+1/4)\alpha^{12/7}\leq0.15$) or $N=1$
and arbitrary $K$
(if $Z\alpha^2\leq0.6$ and $\alpha\leq0.3$) . The problem for general $N$
and $K$ was open for 9 years and we present a surprisingly simple
solution here.

The  bounds in (\ref{eq:4}) on $Z\alpha^2$ and $\alpha$ are not artifacts.
It is shown in \cite{FLL} and \cite{LoY} that the zero-modes cause
$E=-\infty$ when $Z\alpha^2 >11.11$ for the ``hydrogenic" atom, i.e., a
single spin 1/2 particle and one nucleus.  If the number of nuclei is
arbitrary, it is shown in \cite{LL} that there is collapse if
$\alpha>6.67$, no matter how small $Z$ is.  {\it Magnetic stability,
like relativistic stability, implies a (Z-independent) bound on
$\alpha$.}

Prior to our work a proof of type 2 stability for (\ref{eq:1}) with $Z=1$,
$q=2$, and some sufficiently small $\alpha$ was announced
(unpublished) by C.~Fefferman and sketched to one of us.  Our proof is
unrelated to his, considerably simpler and,
more importantly, gives physically realistic constants.

We begin our analysis with the observation that
length scaling considerations suggest that
the key to understanding the stability problem is
somehow to replace ${\cal T}$, on each energy scale, $e$, by
$\mu{\cal T}/e$, where $\mu$ is a fixed energy but $e$ is variable.
On energy scales $e>\mu$ we can use the fact that ${\cal T}>0$ to replace
${\cal T}$ by $\mu{\cal T}/e$ without spoiling lower bounds.
It might seem odd to replace ${\cal T}$ by something smaller,
but what is really happening is that
$\mbox{\boldmath $\sigma$} \cdot\mbox{\boldmath $B$} $ is
being partially controlled by
$[1-\mu e^{-1}](\mbox{\boldmath $p$} +\mbox{\boldmath $A$})^2$.
The idea of replacing ${\cal T}$ by a fraction of ${\cal T}$ was also
used in \cite{FLL}, but no energy dependence was used there.

We shall illustrate this concept by three calculations.  The first, A,
will establish magnetic stability by relating it to the stability of
relativistic matter (see \cite{LY,DL,C,FdL}).  The second, B, will
be the proof of Theorem~2. The
third, C, will use essential parts of the second calculation and an
electrostatic inequality proved in \cite{LY} to prove magnetic
stability without resorting to relativistic stability.

\paragraph*{A. Magnetic Stability from Relativistic Stability:}
We use stability of relativistic matter in the
form proved in \cite{LY}. From the corollary of Theorem 1 in \cite{LY}
with $\beta=0.5$ we have, for any $0<q\kappa \leq 0.032$ and $Z\kappa
\leq 1/\pi$,
\begin{equation}\label{eq:a1}
	\sum_{i=1}^N|\mbox{\boldmath $p$} _i+
	\mbox{\boldmath $A$}_i|+\kappa V_{\rm c} \geq0\ .
\end{equation}
(Although Theorem 1 in \cite{LY} was stated only for
$|\mbox{\boldmath $p$} |$, it holds for
$|\mbox{\boldmath $p$} +\mbox{\boldmath $A$}|$ because
it relies only on the magnitude of the resolvent
which only gets smaller
when $\mbox{\boldmath $A$}$ is not zero. That is ,
$\bigl||\mbox{\boldmath $p$} +\mbox{\boldmath
$A$}|^{-s}(x,y)\bigr|
\leq\bigl||\mbox{\boldmath $p$} |^{-s}(x,y)\bigr|$ for each
$s>0$ and $x,y$ in $\mbox{\boldmath $R$}^3$.
This follows at once from a similar bound on the heat kernel
$\{\exp[-t(\mbox{\boldmath $p$} +\mbox{\boldmath $A$})^2]\}(x,y)$
which, in turn, follows from its representation as a path integral.
This was pointed out in \cite{AHS,CSS}.
Only the resolvent powers
$|\mbox{\boldmath $p$} +\mbox{\boldmath $A$}|^{-s}$
enter the proof of Theorem 1 in \cite{LY}.)

Using (\ref{eq:a1}), $H$ is bounded below by $\bar H = \sum_{i=1}^N h_i$
where $h$ is the one-body operator
$ h = {\cal T} - \kappa^{-1} \vert \mbox{\boldmath $p$}
+\mbox{\boldmath $A$} \vert$.
Thus, $E$  is bounded below by $\varepsilon\int B^2+\bar E_N$,
where $\bar E_N = q\sum_{j=1}^{[N/q]} \varepsilon_j$ and
$\varepsilon_1 \leq \varepsilon_2 \leq...$ are the eigenvalues of
$h$. For $e>0$, let $N_{-e}(h)$ be the number of eigenvalues of
$h$ less than or equal to $-e$. Choose $\mu >0$ and note that
\begin{equation}\label{eq:a3}
        \bar E_N \geq -N\mu -q\int_{\mu}^\infty N_{-e}(h) {\rm d} e\ .
\end{equation}

The crucial step in our proof is noting that the positivity
of the operator
${\cal T}$ implies that ${\cal T}\geq \mu{\cal T}/e$ when $e\geq \mu$.
Thus, ${\cal T}\geq \mu e^{-1}{\cal T}
\geq \mu e^{-1}(\mbox{\boldmath $p$}
+\mbox{\boldmath $A$})^2-\mu e^{-1} B(x)$
when $e\geq \mu$. By Schwarz's inequality,
$\kappa^{-1}|\mbox{\boldmath $p$}
+\mbox{\boldmath $A$}|\leq (1/3)e^{-1}\kappa^{-2}
(\mbox{\boldmath $p$} +\mbox{\boldmath $A$})^2+(3e/4)$ and hence
if we set $\mu=(4/3)\kappa^{-2}$ we obtain
$$
        h\geq e^{-1}\kappa^{-2}(\mbox{\boldmath $p$} +
        \mbox{\boldmath $A$})^2-(4/3)e^{-1}\kappa^{-2}B(x)
        -(3e/4)\equiv h_{e}\ .
$$
Thus, $N_{-e}(h)\leq N_{-e}(h_e)$ and this can be estimated by the
Cwikel-Lieb-Rozenblum (CLR) bound \cite{L2}, i.e.,
$N_{-e}((\mbox{\boldmath $p$} +\mbox{\boldmath $A$})^2-U(x))\leq
L_3\int[U(x)-e]^{3/2}_+ {\rm d}^3x$
where $[a]_+\equiv\max(a,0)$ and $L_3=0.1156$.
In our case:
\begin{equation}\label{eq:a4}
        N_{-e}(h_e)\leq L_3\int\left[\frac{4B(x)}{3}
	-\frac{e^2\kappa^2}{4}
        \right]_+^{3/2}
        {\rm d}^3x\ .
\end{equation}
Inserting this bound in (\ref{eq:a3}), a simple calculation yields
$$
        \bar E_N\geq -N\mu-(2\pi/3) q\kappa^{-1} L_3
        \int B(x)^2{\rm d}^3 x\ .
$$
We choose $\kappa$ so that the field energy terms are non-negative, i.e.,
$\kappa\geq(16\pi^2/3)L_3\alpha^2 q= 6.1\alpha^2q$.
We conclude, by (\ref{eq:a1}), that magnetic stability holds if
\begin{equation}\label{eq:a5}
        q\alpha\leq 0.071\quad\hbox{and}\quad qZ\alpha^2\leq0.052\ .
\end{equation}
For $q=2$, the first condition is $\alpha\leq 1/28$.
For $q=2$ and $\alpha=1/137$ stability occurs if $Z\leq490$.

Assuming (\ref{eq:a5}) holds, we then use (\ref{eq:a1}) and choose $\kappa
= \min\{0.0315q^{-1},
(\pi Z)^{-1}\}$. Our lower bound on the
ground state energy per electron, by
this method, is then $-\mu=-(4/3)\kappa^{-2}=-\max\{1345q^2,
13.2Z^2 \}$.

{\it Remark:\/} We used the CLR bound in (\ref{eq:a4}).
Since the derivation of this bound is not elementary the reader might
wish to use an easier to derive bound---at the cost of worsening the
final constants. A useful substitute  is
$$
        N_{-e}\leq 0.1054
        e^{-1/4}\int\left[U(x)-e/2\right]_+^{7/4}{\rm d}^3 x\ ,
$$
which is in (2.8) of \cite{LT} and which can be derived by means
originally employed for the Lieb-Thirring inequality. This
same remark also applies to our other calculations below.

\paragraph*{B. The Lieb-Thirring Inequality:}
As before we note that $\sum \varepsilon_i=
-\int_0^\infty N_{-e}({\cal T}-U){\rm d}  e$. We write $\int_0^\infty=
\int_0^\mu+\int_\mu^\infty$.
The parameter $\mu$ will be optimized below.
Noting that
${\cal T}\geq (\mbox{\boldmath $p$} +\mbox{\boldmath $A$})^2-B(x)$
and applying the CLR
bound in the same fashion as before to $\int_0^\mu$ yields
\begin{equation}\label{eq:b3}
        L_3\int_0^\mu\int [B(x)+U(x)-e]_+^{3/2}{\rm d}^3x{\rm d}e\ .
\end{equation}

In $\int_\mu^\infty$ we replace ${\cal T}$ by the
lower bound
$\mu e^{-1}[(\mbox{\boldmath $p$} +
\mbox{\boldmath $A$})^2-B(x)]$ and obtain $N_{-e}({\cal T}-U)\leq
N_{-e}(\mu e^{-1}[(\mbox{\boldmath $p$} +
\mbox{\boldmath $A$})^2-B]-U)$. A further
application of the CLR inequality yields the bound on $\int_{\mu}^\infty$
\begin{equation}\label{eq:b4}
        L_3\int_{\mu}^\infty\int[B(x)+(e/\mu)U(x)-
        e^2/\mu]_+^{3/2}{\rm d}  x{\rm d}  e\ .
\end{equation}
It is easy to see that for any $0<\gamma<1$ the integrand in (\ref{eq:b3})
is bounded
above by
$$
        \sqrt{2}\left([B(x)-\gamma e^2/\mu]_+^{3/2}+
        [U(x)-(1-\gamma)e]_+^{3/2}\right)\ .
$$
Treating the integrand in (\ref{eq:b4}) in a similar
fashion and combining the inequalities we find
\begin{eqnarray*}
        \sum|\varepsilon_i|&&\leq\sqrt{2}L_3
        \int\biggl\{\int_0^\infty [B(x)-
        \gamma e^2/\mu]_+^{3/2}{\rm d}  e \nonumber\\&&
        +\int_0^\mu [U(x)-(1-\gamma) e]_+^{3/2}{\rm d} e
        \\ &&
        +\int_\mu^\infty\left[(e/\mu) U(x) -
        (1-\gamma) e^2/\mu\right]_+^{3/2}{\rm d} e
        \biggr\}{\rm d}^3x\ .\nonumber
\end{eqnarray*}
After extending the last two integrals to $\int_0^\infty$, a straightforward
computation yields
\begin{eqnarray*}
        \sum|\varepsilon_i|\leq&&\sqrt{2}L_3\int
	\biggl\{\frac{2}{5(1-\gamma)} U(x)^{5/2}
        +\frac{3\pi\mu^{1/2}}{16\gamma^{1/2}} B(x)^2\\ &&
\mbox{} + \frac{3\pi}{128}
        \mu^{-3/2}(1-\gamma)^{-5/2} U(x)^4\biggr\}{\rm d}^3x\ .
\end{eqnarray*}
Optimizing over $\mu$ yields (\ref{eq:lt2}).

To prove the more general form of (\ref{eq:lt2}) replace $\mu e^{-1}$
         by $(\mu e^{-1})^s$, where $s=2p/3 - 5/3$.

\paragraph*{C. Proof of Theorem 1:}
We turn now to our third illustration of the
concept of running energy scale and
prove the stability directly, not relating it to the relativistic problem.
By this method we get the correct dependence of the ground state
energy on Z and also somewhat better critical constants than in (\ref{eq:a5}).

Following \cite{LY} we first replace the Coulomb potential
by a single particle
potential in (\ref{eq:c1}) below. We break up $\mbox{\boldmath $R$}^3$
into Voronoi cells defined by the nuclear
locations, i.e., $\Gamma_j = \{ x: |x-R_j| \le |x-R_k|\
{\rm for\  all}\  k \}$ is the
j--th Voronoi cell. Each $\Gamma_j $ contains a ball
centered at $R_j$ with
radius $D_j=$ min$\{|R_j-R_k|:j \ne k\}/2$.

The following bound on $V_{\rm c}$ is proved in \cite{LY}:
Choose some $0<\lambda<1$. Then
\begin{equation}\label{eq:c1}
        V_{\rm c}\ge -
        \sum_{i=1}^N W(x_i)+
        \sum_{j=1}^K \frac{Z^2}{8D_j} \ ,
\end{equation}
where $W(x)=Z|x-R_j|^{-1}+F_j(x)$ for $x \in \Gamma_j$ with
$F_j(x)$ defined by
\begin{eqnarray*}
(2D_j)^{-1}(1-D_j^{-2}|x-R_j|^2)^{-1}&\ \ \mbox{for }\ &
|x-R_j| \le \lambda D_j\\
(\sqrt {2Z}+1/2)|x-R_j|^{-1} &\ \ \mbox{for }\ &
|x-R_j| > \lambda D_j\ .
\end{eqnarray*}
The point about this inequality is that the potential
$W$ has the same singularity
near each nucleus as $V_c$ has, and that
the rightmost term in (\ref{eq:c1}) is repulsive.
This term will be responsible for stabilizing the system.

The problem is thus reduced to obtaining a lower bound on
$q\sum^\prime \varepsilon_j$, where
$\sum^\prime \varepsilon_j$ is the sum of the first $[N/q]$
negative eigenvalues of
of ${\cal T} - W$. Note that Theorem~2
cannot be applied directly
to this problem, since $W$ is
neither integrable to the power $5/2$ nor to the power 4.
Instead we have to do the calculations directly.

For $\nu >0$ (a number that is chosen later) set $W_{\nu}(x)\equiv
(W(x)-\nu)_+$ and note that $W(x)-\nu\leq W_\nu(x)$. Then, as in
(\ref{eq:a3}), $ q\sum^\prime \varepsilon_j \ge-N\nu  -q \int_0^\infty
N_{-e}({\cal T}- W_{\nu}) {\rm d}  e $. Again,
\begin{eqnarray}
        \int_0^\infty N_{-e}({\cal T}- W_{\nu}) {\rm d}  e &\le&
        \int_0^{\mu} N_{-e}({\cal T}- W_{\nu}) {\rm d}  e \nonumber\\
        &+&\int_{\mu}^\infty N_{-e}(\mu e^{-1}{\cal T}- W) {\rm d} e\ ,
	\label{eq:c2}
\end{eqnarray}
where we have replaced  $W_{\nu}(x)$ by  $W(x)$ in
the second term. Applying  the CLR bound to the first
expression on the right
side we obtain
$ L_3 \int \int_0^{\mu} [B(x)+
W_{\nu}(x)-e]^{3/2}_+{\rm d} e {\rm d} ^3x $
which can be bounded, as in part B, by
\begin{eqnarray}
        \sqrt 2 L_3 \int \biggl\{ && \int_0^\mu[B(x)-
	\frac{\gamma e^2}{\mu}]^{3/2}_+{\rm d}  e \nonumber\\ && +\mbox{}
        \frac{2}{5}(1-\gamma)^{-1}W_{\nu}(x)^{5/2} \biggl\}{\rm d} ^3x
        \ , \label{eq:c3}
\end{eqnarray}
for any $0< \gamma <1$.

The difficulty in dominating the second term in (\ref{eq:c2}) comes
from the Coulomb
singularity of $W(x)$ which is not fourth power integrable.
The singularity can be controlled using the following operator inequality
which follows from the diamagnetic inequality
$\int |(\mbox{\boldmath $p$}+\mbox{\boldmath $A$})\psi|^2 {\rm d} ^3x \ge
\int \bigl|\mbox{\boldmath $p$}|\psi|\bigr|^2 {\rm d} ^3x$
and Lemma~2a on p.~708 of \cite{LD}.
$$
        (\mbox{\boldmath $p$} +\mbox{\boldmath $A$})^2-Z/|x|\geq
        -\cases{Z^2/4 +(3/2)ZR^{-1},& if $|x|\leq R$\cr
                Z|x|^{-1},& if $|x|\geq R$}\ .
$$
Choose $R=\lambda D_j$ and write $(\mbox{\boldmath $p$} +
\mbox{\boldmath $A$})^2
=\beta(\mbox{\boldmath $p$} +\mbox{\boldmath $A$})^2+
(1-\beta)(\mbox{\boldmath $p$} +\mbox{\boldmath $A$})^2$
for some $0<\beta<1$. Then, by scaling,
$$
        (\mu/e){\cal T} -W_{\nu}\ge (\mu/e)(1-\beta)
        (\mbox{\boldmath $p$}+\mbox{\boldmath $A$})^2-(\mu/e)B
        -\widetilde W\ ,
$$
where $\widetilde W(x,e)= \widetilde G_j(x,e)+F_j(x)$
for $x \in \Gamma_j$ with $\widetilde G_j(x,e)$ defined by
\begin{eqnarray*}
        (Z^2e/4\beta\mu) + 3Z/(2\lambda D_j) &\ \  \mbox{for }\ &
        |x-R_j| \le \lambda D_j\\
        Z|x-R_j|^{-1} &\ \ \mbox{for }\ & |x-R_j| > \lambda D_j\ .
\end{eqnarray*}
Note that $\widetilde W$ depends on $e$.

Again, as in part B, we can use the CLR bound on the second
term in (\ref{eq:c2}) to obtain
[when $1-\gamma\geq Z^2/(4\beta\mu)$]
\begin{eqnarray}
        \sqrt 2&&L_3(1-\beta)^{-3/2}
        \int\biggl\{\int_{\mu}^{\infty}[B(x)-\gamma e^2/\mu]_+^{3/2}
        {\rm d}  e \nonumber\\ &&+ {\mu}^{-3/2}
        \int_0^\infty[e \widetilde  W(x,e)
        -(1-\gamma)e^2]_+^{3/2}{\rm d}  e \biggr\}{\rm d}  ^3x \ .
	\label{eq:c4}
\end{eqnarray}
First we compute the last integral in (\ref{eq:c4}), which is
$$
        \sum_{j=1}^K\int_{\Gamma_j}\int_0^\infty [e \widetilde
	G_j(x,e)+e F_j(x)-(1-\gamma)e^2]_+^{3/2}{\rm d} e
        {\rm d}^3x\ .
$$
Now split the $\Gamma_j$ integral into an inner integral
$|x-R_j| \le \lambda D_j$
and an outer integral $|x-R_j| > \lambda D_j$.
The inner integral yields, using the definitions of
$\widetilde G_j$ and $F_j$,
\begin{eqnarray}
        \frac{3\pi^2}{32}(1-\gamma-\frac{Z^2}{4\beta \mu})^{-5/2}
	\int_0^{\lambda}\biggl[&&\frac{1}{2(1-r^2)}\nonumber\\ &&
	\mbox{}+\frac{3Z}{2\lambda}\biggr]^4r^2{\rm d} r
        D_j^{-1}\ . \label{eq:c6}
\end{eqnarray}
To bound the outer integral from above we replace $\Gamma_j$
by $\mbox{\boldmath $R$}^3$ and get
\begin{equation}
        (3 \pi^2/32)(1-\gamma)^{-5/2}(\sqrt Z +\sqrt{1/2})^8
        (\lambda D_j)^{-1}\ . \label{eq:c7}
\end{equation}
Combining (\ref{eq:c3})--(\ref{eq:c7}) we find that the sum of the
negative eigenvalues of
${\cal T} - W_{\nu}$ is bounded below by
\begin{equation}
        -a\int W_{\nu}(x)^{5/2}{\rm d}^3x -b\int B(x)^2 {\rm d}^3x
        -c\sum_{j=1}^K D_j^{-1} \ .
        \label{eq:c8}
\end{equation}
Here $a=q(2 \sqrt 2/5)L_3(1-\gamma)^{-1}$,
\begin{eqnarray*}
        b&=&q\frac{3\pi \sqrt 2}{16}L_3(1-\beta)^{-3/2}
        (\mu/\gamma)^{1/2}\\
        c&=&q\frac{3\pi^2 \sqrt 2}{32}L_3(1-\beta)^{-3/2}
        \mu^{-3/2}
        \Bigl\{
	\frac{(\sqrt Z +\sqrt{q/4})^8}{\lambda(1-\gamma)^{5/2}}
	\\ &&+
        (1-\gamma-\frac{Z^2}{4\beta \mu})^{-5/2} \int_0^{\lambda}
        \Bigl[\frac{q}{4(1-r^2)}+\frac{3Z}{2\lambda}\Bigr]^4
	r^2{\rm d} r \Bigr\}.
\end{eqnarray*}
To simplify the stability condition we have artificially increased the
bounds by recalling that $q\geq 2$ and twice replacing $1/2$ by
$q/4$ in the
definition of $c$.  We choose $\beta=1/8$, $\gamma=1/2$, $\lambda=8/9$
and $\mu$ so that $b=(8\pi\alpha^2)^{-1}$. The {\it stability condition}
$c\leq Z^2/8$ [see (\ref{eq:c1})] now depends {\it only} on the 2
parameters $X=qZ\alpha^2$ and
$Y=q\alpha$. A straightforward, but lengthy calculation shows that the
stability condition holds if $X=X_0\equiv0.082$ and $Y=Y_0\equiv0.12$.
The condition is monotone in $Y$, so
it holds for $X=X_0$, $Y\leq Y_0$. Although our condition does not hold
for {\it all} $X\leq X_0$, $Y\leq Y_0$ we can use the $Z$-monotonicity
of $E$ to conclude stability in this range; this proves (\ref{eq:4}).
With the same values of $\beta,\gamma$ and
$\lambda$ and with $q=2$ the values $Z=1050$, $\alpha=1/137$ also
give stability.

To derive (\ref{eq:3}), note that
$W(x)\leq Q|x-R_j|^{-1}$ for $x\in\Gamma_j$.
Using this bound  and replacing
$\Gamma_j$ by $\mbox{\boldmath $R$}^3$, one easily obtains
$-\sqrt{2}\pi^2L_3qKQ^3\nu^{-1/2}-N\nu$ as a lower bound
on the $-a\int W_\nu^{5/2}$ term in (\ref{eq:c8}).
Optimizing over $\nu$ yields (\ref{eq:3})
when $X=X_0$, $Y\leq Y_0$.
In this case, $Z\geq Z_0\equiv 5.7 q$. If $X\leq X_0$, $Y\leq Y_0$
and $Z\geq Z_0$ we get a lower bound on $E$ by increasing $\alpha$
until $X=X_0$, $Y\leq Y_0$;
this yields (\ref{eq:3}) with $Q=Q(Z)$. Otherwise, with
$Z<Z_0$, we use the $Z$-monotonicity of $E$ to conclude (\ref{eq:3})
with $Q=Q(5.7 q)$.

This work was partially supported by NSF grants PHY90--19433-A04 (E.H.L.),
DMS92--07703 (M.L.) and DMS92--03829 (J.P.S.).

\end{document}